\documentclass[10pt, conference, letterpaper]{IEEEtran}

\IEEEoverridecommandlockouts
\usepackage{cite}
\usepackage{amsmath,amssymb,amsfonts}
\usepackage{algorithmic}
\usepackage{graphicx}
\usepackage{textcomp}
\usepackage{xcolor}
\usepackage{booktabs}
\usepackage{diagbox}
\usepackage{multirow}
\usepackage{subfigure}
\usepackage[justification=centering]{caption}
\usepackage{enumitem}
\usepackage{pifont}

\def\BibTeX{{\rm B\kern-.05em{\sc i\kern-.025em b}\kern-.08em
    T\kern-.1667em\lower.7ex\hbox{E}\kern-.125emX}}

\newcommand{\tb}{\noindent\textbf}
\newcommand{\eb}{\noindent\emph}

\newcommand{\ie}{\emph{i.e.}}
\newcommand{\eg}{\emph{e.g.}}

\begin{document}


\title{
BiSwift: Bandwidth Orchestrator for Multi-Stream Video Analytics on Edge}


\author{
\IEEEauthorblockN{Lin Sun\IEEEauthorrefmark{1}\IEEEauthorrefmark{5},
Weijun Wang\IEEEauthorrefmark{2}\IEEEauthorrefmark{5}, 
Tingting Yuan\IEEEauthorrefmark{3}\IEEEauthorrefmark{4},
Liang Mi\IEEEauthorrefmark{1}, 
Haipeng Dai\IEEEauthorrefmark{1}\IEEEauthorrefmark{4}, 
Yunxin Liu\IEEEauthorrefmark{2},
Xiaoming Fu\IEEEauthorrefmark{3}
}

\IEEEauthorblockA{
\IEEEauthorrefmark{1}State Key Laboratory for Novel Software Technology, Nanjing University, China\\
\IEEEauthorrefmark{2}Institute for AI Industry Research (AIR), Tsinghua University, China\\
\IEEEauthorrefmark{3}Institute of Informatik, University of G{\"o}ttingen, Germany\\
\IEEEauthorrefmark{5}Co-first authors 
\IEEEauthorrefmark{4}Corresponding authors
}}

\maketitle

\begin{abstract}
High-definition (HD) cameras for surveillance and road traffic have experienced tremendous growth, demanding intensive computation resources for real-time analytics. 
Recently, offloading frames from the front-end device to the back-end edge server has shown great promise. 
In multi-stream competitive environments, efficient bandwidth 
management and proper scheduling are crucial to ensure both high inference accuracy and high throughput.
To achieve this goal, we propose BiSwift, a bi-level framework that scales the concurrent real-time video analytics by a novel adaptive hybrid codec integrated with multi-level pipelines, and a global bandwidth controller for multiple video streams.
%
The lower-level front-back-end collaborative mechanism (called adaptive hybrid codec) locally optimizes the accuracy and accelerates end-to-end video analytics for a single stream. 
The upper-level scheduler aims to accuracy fairness among multiple streams via the global bandwidth controller.
The evaluation of BiSwift shows that BiSwift is able to real-time object detection on 9 streams with an edge device only equipped with an NVIDIA RTX3070 (8G) GPU.
BiSwift improves 10\%$\sim$21\% accuracy and
presents 1.2$\sim$9$\times$ 
throughput compared with the state-of-the-art video analytics pipelines.

\end{abstract}

\section{Introduction}

\begin{figure}[t!]
\centering
\includegraphics[width=0.47\textwidth]{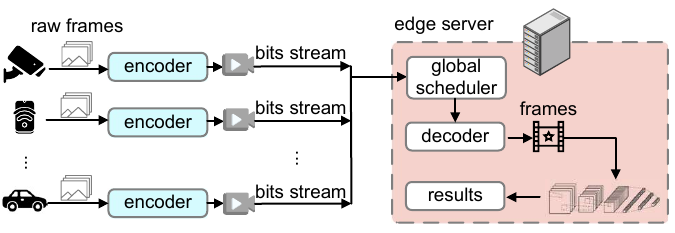}
\caption{Multi-stream video analytics on edge server.}
\label{fig:intro}
\vspace{-2em}
\end{figure}

Autonomous analytics of videos generated by ubiquitous video cameras have great potential with advances in computer vision. 
Although Deep Neural Networks (DNNs) \cite{bochkovskiy2020yolov4, ahn2018fast, ren2015faster, wang2021swiftnet} approaches have been developed to dramatically improve the accuracy of various vision tasks, they introduce stringent demands on computing resources.
As a result, videos need to be streamed to the remote edge server for inference in video analytics pipelines (VAP) \cite{du2020server, li2020reducto, padmanabhan2022gemel}, as illustrated in Fig. \ref{fig:intro}, due to the limited computing resources in the front devices.

This distributed architecture shifts the burden to the network.
Accurate video analytics stands on the informative input, but high-quality streaming results in expensive network bandwidth costs.
For example, the inference accuracy of Faster R-CNN \cite{ren2015faster}, a DNN-based object detector, increases from 61\% to 92\% when changing the input video from 540P to 1080P, but it requires 3.7$\times$ network bandwidth \cite{Yoda_c}. 

Today's VAPs hardly offer enough video quality for multiple streams due to the limited networking bandwidth. 
This fact could surprise people watching 2K, 4K, or even 8K videos on Netflix and YouTube.
However, the uplink bandwidth from the front-end camera to the edge server is up to ten times lower than the downlink from the content server to users for video on demand \cite{speedtest}.
For example, the limited available network bandwidth of Wi-Fi LAN (Local Area Network) is only set to 15Mbps for multiple video cameras in \cite{cell}. Some prior work \cite{wang2020joint,meiling2021blockchain,peng2021secure,lingshu2021secure} focuses on addressing network security, reliable services, dynamic management or efficient scheduling.





To address the issue of limited bandwidth, prior studies have proposed various approaches, mainly classified into two categories: front-end compression and back-end enhancement.
Front-end compression utilizes lightweight DNNs or edge feedback information to filter out invalid frames (\eg, frames containing empty street in traffic video)  \cite{hsieh2018focus, zhang2015design, Kang2017NoScope, li2020reducto} or fine-grained invalid content (pixels) within frames (\eg, the street part is invalid compared with vehicles)
\cite{du2020server,  liu2019edge, xie2019source, du2022accMPEG}. This approach delivers only valid parts of the video, resulting in significant bandwidth savings. 
While back-end enhancement on edge decodes compressed video into frames and applies super-resolution (SR) to improve video quality
\cite{ yeo2020nemo, kim2020neural, yeo2018neural, yi2020supremo, yeo2022neuroscaler,yeo2017will} or analytical accuracy \cite{yi2020eagleeye, AccDecoder}; it also provides considerable bandwidth saving via bypassing HD streaming over the uplink.

\begin{figure*}[t]
\setlength{\abovecaptionskip}{-1pt}
\setlength{\belowcaptionskip}{-9pt}
\begin{center}
\subfigure[Bandwidth utilization of various streaming protocols, of which \emph{adaptive qp} that often used in video analytics systems utilizes bandwidth more aggressively than WebRTC.
] {
 \label{fig:BWutilization}     
\includegraphics[width=0.23\textwidth]{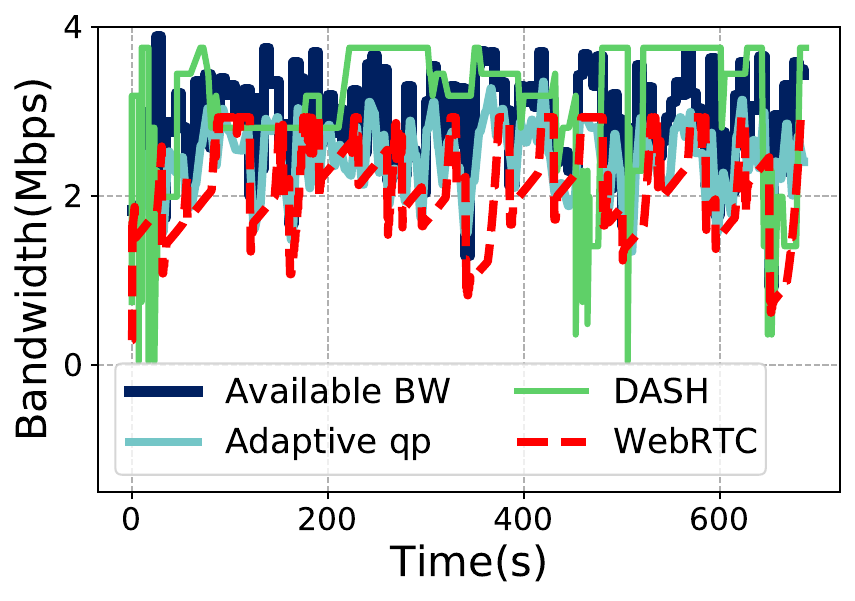}
}
\hspace{0.1em}
\subfigure[Accuracy and size (Top), and latency (Bottom) against an image quality factor from 0 to 100, showing the high accuracy and low latency in a quality range 40$\sim$80.] {
\label{fig:encode_params} 
\includegraphics[width=0.23\textwidth]{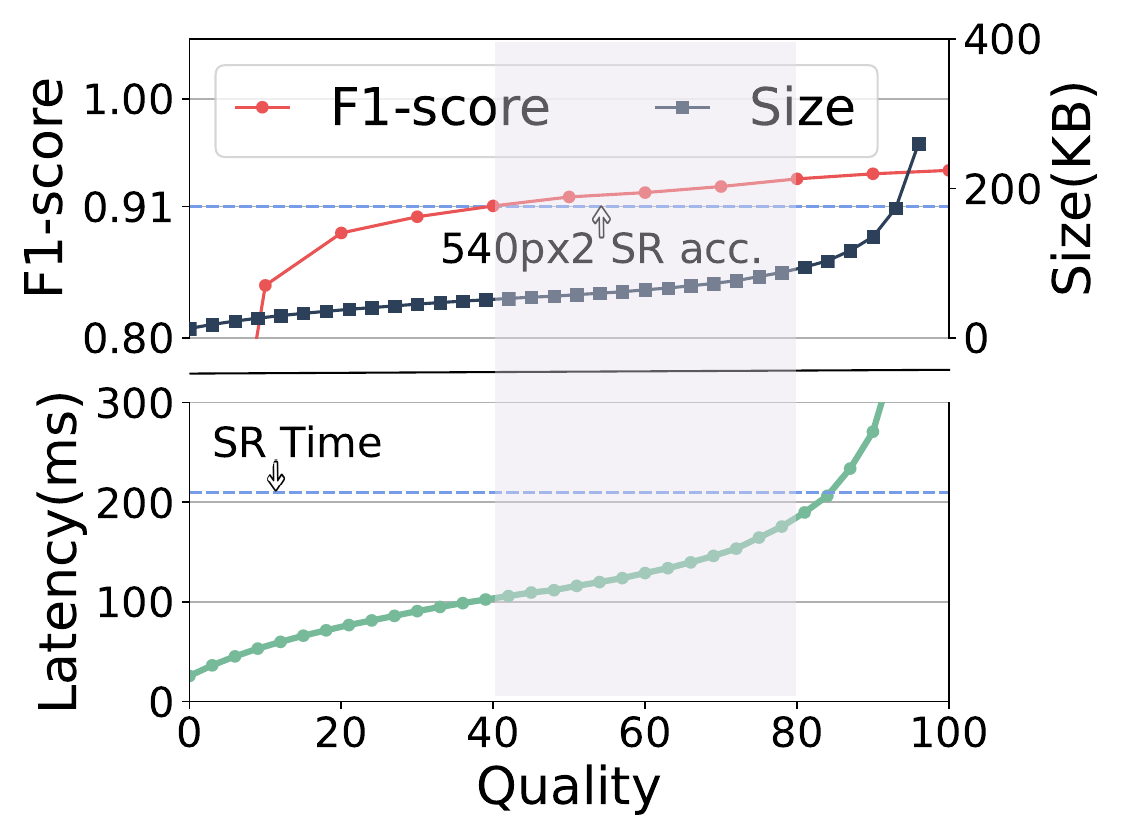}


}
\hspace{0.1em}
\subfigure[Accuracy of HD compared with neural-enhanced 
images with various input size and scaling factors causing $>8.1\%$ accuracy degradation.] { 
\label{fig:f1_diff}
\includegraphics[width=0.21\textwidth]{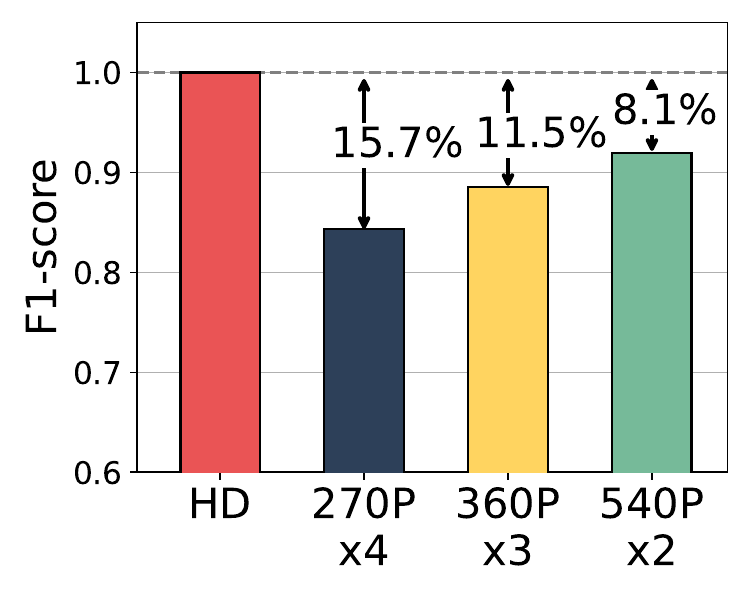}
}
\hspace{0.1em}
\subfigure[Different content in streams implies various robustness of resolution. Stream 1 with large and sparse objects exhibits robustness against degraded input video.] {
\label{fig:band_alloc}
\includegraphics[width=0.23\textwidth]{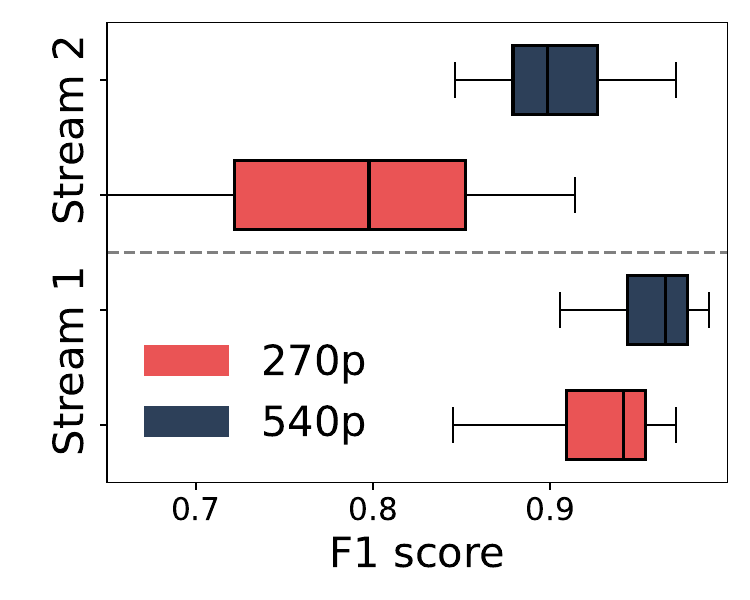}

}
\caption{Motivation of BiSwift.}
\end{center}
\label{fig:motivation}
\vspace{-1.2em}
\end{figure*}

While these approaches have provided significant benefits, we contend that they may lack sufficient robustness under multi-stream scenarios.
Their primary focus on bandwidth saving neglects the exploration of strategies to fully utilize available resources.
We observe that the state-of-the-art streaming protocol in video analytics does not fully leverage the available bandwidth (detailed in Fig. \ref{fig:BWutilization} of \S \ref{sec:motivation}). This limitation may be more evident in multi-stream scenarios.

In this paper, we demonstrate that multi-stream competing video analytics is accurate and effective 
with careful bandwidth management.
To this end, we present BiSwift, a novel video streaming framework that combines hybrid codec and bandwidth orchestration to optimize fairness among multiple streams.
BiSwift allocates stream bandwidth by holistic considering the 
video content, inference accuracy, and server payload. 
It efficiently delivers a few selected set of high-quality images and highly compressed video to the edge, while also leveraging the high-quality images to benefit the compressed video on the edge, thereby improving the inference accuracy.
The resulting system effectively increases throughput (\ie, number of streams) by a factor of 1.2$\sim$9$\times$.

BiSwift consists of two innovative components: the hybrid codec, which includes an encoder and a decoder, and a bandwidth controller.
At the cameras, the hybrid encoder (\S \ref{sec:hybridenc}) first classifies raw images and then encodes the \emph{anchor} type of frames that provide a significant benefit to accuracy improvement into high-definition (HD) images. These HD images, along with the highly compressed video with all encoded frames, are then delivered to the edge server. 
The hybrid decoder (\S \ref{sec:hybriddec}) decodes HD images and video into frames, then feeds them into corresponding pipelines for inference at various levels of accuracy and speed.
The high-quality content from HD images is transferred to compressed video through a quality transfer mechanism (more details in \S \ref{sec:hybriddec}), thereby enhancing the quality of video frames.
At the edge, the bandwidth controller (\S \ref{sec:BWcontroller}) optimizes overall inference performance, by globally balancing the bandwidth allocation for each stream based on their respective demands. It ensures fairness among multiple streams while maximizing the utilization of available bandwidth.
\begin{figure}
    \centering    \includegraphics[width=0.28\textwidth]{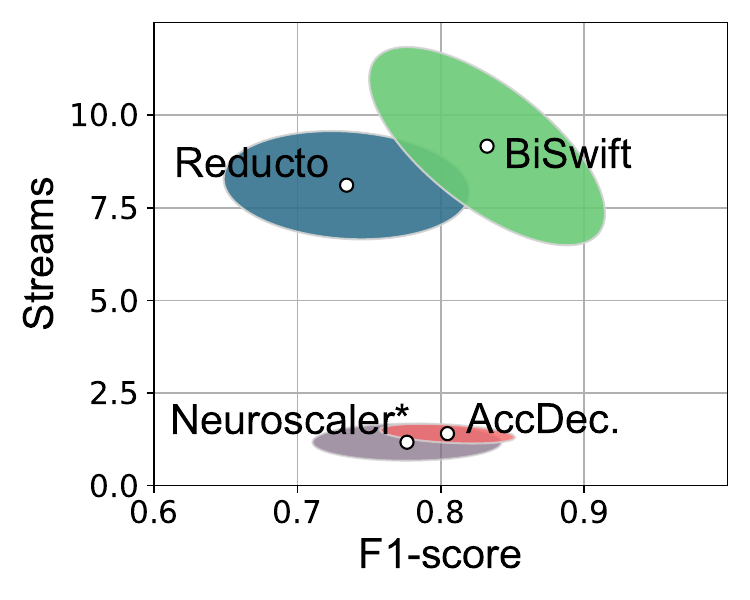}
    \vspace{-0.5em}
    \caption{BiSwift vs. baselines.}
    \label{fig:moti_baseline}
    \vspace{-2.0em}
\end{figure}

\tb{Challenges.}
\label{challenge}
1) To deal with a single stream, 
BiSwift employs adaptive bandwidth allocation to handle the streaming of HD images and video.
A camera transmits HD images to the edge, sharing bandwidth with the video stream. 
Allocating a large bandwidth to HD images can improve their inference accuracy. 
However, this approach leaves less bandwidth for video, which may result in significant accuracy degradation, even when applying quality transferring techniques.
2) Given the bandwidth allocated to HD images, the performance of the HD image stream is also influenced by anchor selection. 
BiSwift addresses this by developing a customized metric for frame classification that considers factors such as video content and the allocated bandwidth, ensuring the most relevant and informative frames are selected as anchors to maximize accuracy improvement. 
3) BiSwift needs to manage bandwidth to optimize overall performance across multiple competing streams.
Unreasonable bandwidth allocation not only limits the value of streams being delivered, but also results in bandwidth underutilization when streams are too simplistic (\eg, a stream with only empty streets for traffic flow analysis).

\tb{Approach.} To address these issues, we first model the problem as a bi-level optimization problem. The high-level handles the allocation of  available bandwidth among all streams, while  the low-level is responsible for deciding the bandwidth ratio for transmitting the video chunk compared to images and performing anchor selection to accelerate single-stream video analytics.
We take advantage of deep reinforcement learning (DRL) to obtain several critical parameters, thus solving the above bi-level optimization issue and achieving a good trade-off between accuracy and latency under high throughput.\\
\tb{Contributions}  are summarized as follows.
\begin{itemize}[left=0.2em]
\setlength{\itemsep}{0pt}
\setlength{\parsep}{0pt}
\setlength{\parskip}{0pt}
\item To the best of our knowledge, BiSwift is the first system to enable real-time object detection on up to 9 video streams with an edge device only equipped with an NVIDIA RTX3070 (8G) GPU under various bandwidth constraints.
    It improves 
    10\%$\sim$21\% accuracy and presents 1.2$\sim$9$\times$ throughput compared with the state-of-the-art VAPs (see Fig. \ref{fig:moti_baseline}).
    \item Our hybrid codec framework, integrated with a multi-level pipeline executor, is shown to deliver a 
    19\% accuracy gain compared with the original pure video codec.
    \item We show BiSwift's bandwidth controller improves accuracy by 8\%
    and bandwidth utilization by 52\%  
    compared with traditional even bandwidth allocation.
\end{itemize}

\vspace{-1em}

\section{Background}\label{sec:background}
\setlength{\itemsep}{0pt}
\setlength{\parsep}{0pt}
\setlength{\parskip}{0pt}
\vspace{-0.5em}

\tb{Video delivery protocol} is crucial in today's video analytics.
Accurate analytics requires compute-intensive DNNs that cheap video cameras cannot afford 
\cite{padmanabhan2022gemel}, thus demanding a \emph{distributed} architecture.
However, such a distributed system lacks a standard delivery protocol. 
Prior studies (\eg, \cite{du2020server, zhang2018awstream, jiang2018chameleon}) have focused mainly on adjusting video encoding parameters of 2-4 second chunks over TCP and UDP to adapt to available bandwidth.
WebRTC \cite{WebRTC} is a popular standard for live video ingest, ensuring high-quality video with low latency. 
Its client side uses Google Congestion Controller (GCC) \cite{GCC} over the Real-Time Protocol (RTP) \cite{RTP} for adaptive bitrate streaming, dynamically encoding the video to match the available network bandwidth.
Dynamic Adaptive Streaming over HTTP (DASH) \cite{stockhammer2011dynamic} is a widely accepted streaming protocol for video on demand. 
It pre-encodes video chunks into multiple bitrate representations on delivery servers, enabling clients to adaptively switch between them based on network bandwidth.

\tb{Neural-enhanced video analytics} \cite{AccDecoder, yi2020eagleeye, wang2022enabling} and video playback \cite{yeo2020nemo, yeo2022neuroscaler, kim2020neural, yeo2018neural, yi2020supremo} leverage super-resolution (SR) DNNs to enhance low-resolution images/frames to high-definition ones on an edge server, improving accuracy and quality.
Pre-trained SR models learn a pixel generator mapping neighboring pixels' values to the generated one from training video data.
These models are then deployed in runtime systems to generate high-definition content in real-world videos.

Previous experiments have shown that the neural-enhanced approach can save up to 21\% network bandwidth \cite{yeo2018neural}, increase the quality of experience by 31.2 \cite{yeo2020nemo}, and improve accuracy by up to 21\% accuracy \cite{AccDecoder}, but with higher computational costs.

\tb{Selective neural-enhancement} reduces the computing cost by using temporal redundancy across video frames \cite{zhang2017fast,yeo2020nemo,yeo2022neuroscaler}.
A few frames are enhanced by the SR model, while the remaining frames are efficiently up-scaled by reusing SRed ones via codec information.
However, such reuse leads to image quality loss accumulating across consecutive non-SRed frames.
Prior work \cite{yeo2020nemo, yeo2022neuroscaler} proposed frame selection approaches to relieve loss accumulation by resetting in selected frames.
However, their centralized workflow, without considering network resources, cannot afford our scenario.

\section{Motivations}\label{sec:motivation}
\tb{Limited room for improvement of video delivery.} 
Video streaming in the analytics pipeline is sensitive to the variation in available network bandwidth.
Fig. \ref{fig:BWutilization} shows the respective video bitrate selection results of 1) the WebRTC protocol, 2) the DASH protocol, and 3) the widely used adaptive quantization parameter (QP) mechanism in video analytics pipelines \cite{zhang2018awstream, jiang2018chameleon, du2020server, du2022accMPEG}, within the available bandwidth given by an FCC broadband network trace \cite{FCC}.
Compared with DASH, WebRTC and adaptive qp use bandwidth much more conservatively because 1) they must avoid packet loss to minimize latency due to retransmission; 2) the edge server cannot use much buffer to absorb the bandwidth variations in live video scenario; 3) they cannot pre-encode various-bitrate version, as DASH does, to provide the chance to probe and retransmit the same-content but different-bitrate video.
Compared with WebRTC, adaptive qp uses bandwidth more aggressively by sharply adjusting the encoding quality because it does not need to smooth the quality of successive chunks like WebRTC does for a better quality of user experience.
As a result, from the perspective of delivery protocol, the bandwidth utilization of video analytics pipeline makes sounds,  
we argue that traditional codec designed for human perception is not appropriate for video analytics.

\tb{Insight \#1.} \emph{A novel hybrid codec compatible with the adaptive qp or bitrate mechanism of video analytics is promising.}

\tb{Drawbacks of neural-enhanced video analytics.} 
First, the execution of the super-resolution is computationally intensive and time consuming, becoming worse in multi-stream environments.
Because GPU commonly has limited memory size (\eg, 8G in NVIDIA RTX3070), multiple content-aware DNNs specified for different streams \cite{yeo2022neuroscaler}, each requiring several GBs of memory, must be frequently swapped in/out, causing nonnegligible overhead (135ms for EDSR \cite{Edsr_w}).
Besides, recent work \cite{AccDecoder} only selects anchors (7\%$\sim$8\% of the frames) executing neural enhancement for trade-off of accuracy and latency. 
Yet, they overlook the fact that if only a few HD frames are needed to gain benefits across the entire video, why not directly transmit the corresponding HD frames to the server? 
Fig. \ref{fig:encode_params} shows that HD images encoded by JPEG \cite{jpeg}  have high inference accuracy and acceptable size and transmitting latency when the quality factor\footnote{JPEG codec defines quantization values for different quality factors.} is properly set.
Fig. \ref{fig:f1_diff} demonstrates the accuracy gain of transmitting HD images compared with compressing them into video and then upscaling to the original high resolution by SR models. 

\tb{Insight \#2.} \emph{HD image delivery for higher accuracy is promising, provided the anchors are under proper selection.
}



\tb{Inefficient analytics-agnostic bandwidth allocation.}
\label{moti_3}
We observed that total bandwidth must be allocated properly across streams driven by DNN inference accuracy  to maximize the overall accuracy improvement.
As shown in Fig. \ref{fig:band_alloc}, the video full of dense small objects (\ie, Stream 2 with the number of objects over 30, but the average size of each object smaller than 1\% of the frame size) are susceptible to compromised input data, while the video with large objects (\ie, Stream 1 with the number of objects less than 5, and the average size of each object larger than 10\% of the frame size) tends to maintain greater robustness against degraded input data. 
As a result, Stream 2, which exhibits poor robustness with decreasing resolution, should be allocated sufficient bandwidth to avoid its low resolution\footnote{
High-resolution videos are often encoded under high bitrate, occupying a larger bandwidth for transmitting.
}. 
\tb{Insight \#3.} \emph{It becomes paramount to judiciously allocate bandwidth across multiple streams by an analytics-aware method, considering the heterogeneous streams.}

\section{BiSwift}
\begin{figure}[h!]
\centering
    \includegraphics[width=0.49\textwidth]{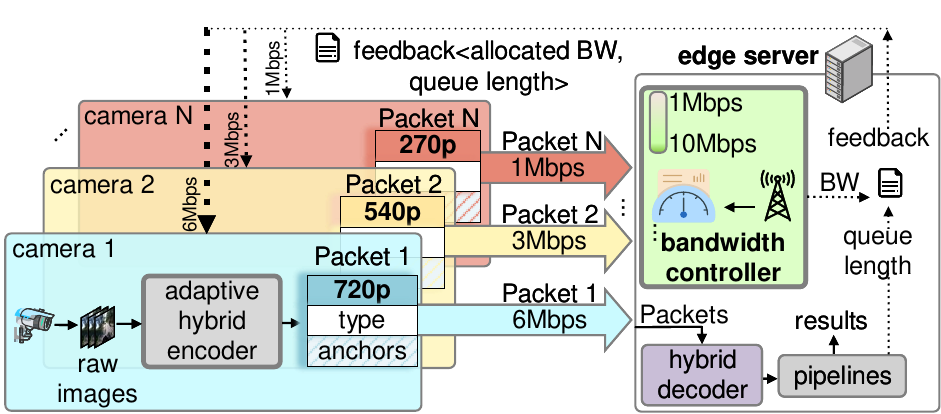}
    \caption{Architecture of BiSwift.}
    \label{fig:BiSwift}
    \vspace{-1.5em}
\end{figure}

%
Motivated by the above insights, BiSwift boosts the accuracy and latency performance of individual streams by utilizing HD images, simultaneously, and optimizes the accuracy fairness\footnote{BiSwift is flexible to any customized objective functions, this paper takes the accuracy fairness as an example to verify the feasibility of BiSwift.} among multiple video streams under limited network bandwidth.
Fig. \ref{fig:BiSwift} shows the overall workflow of BiSwift.
When raw images are captured at a source (\eg, cameras), BiSwift applies a hybrid encoder (see details in \S \ref{sec:hybridenc}) to the raw images by adaptively encoding frames into LR (Low Resolution) video based on the given bandwidth and latency constraint,  and selecting and encoding a few frames (as anchors) into HD images based on the feedback from server, image content and codec-level information (\eg, residual and frame reference).
The resulting encoded packets are delivered to an edge server.
The hybrid decoder (see details in \S \ref{sec:hybriddec}) on edge then decodes the packets and sends them to the respective pipelines for further processing.
The bandwidth controller on the edge server allocates network resources to each stream (every 10 seconds by our empirical setup) as part of the feedback provided to the cameras. 

\subsection{Key Component: Hybrid Encoder}\label{sec:hybridenc}
\tb{Goal.}
Our goal is to maximize the accuracy of each stream while ensuring fairness when processing a large number of streams. 
As for one stream, the frame-grained importance to overall inference accuracy is heterogeneous and the available resources change over time.
In this component, we focus on local decision-making, that is, how to distinguish important frames 
and set up their encoding parameters 
to maximize accuracy within a given bandwidth budget.
We leave the cross-stream decision-making to bandwidth controller \S \ref{sec:BWcontroller}.


Previous neural-enhanced analytical methods \cite{AccDecoder, yeo2022neuroscaler}, which deliver LR video to the backend for super-resolution enhancement, miss two key opportunities for maximizing accuracy.
First, the camera can naturally access the most informative raw images to help make the best decisions.
Second, neural-enhanced methods lead to suboptimal performance in both accuracy and latency. 
Our study in Fig. \ref{fig:encode_params} shows that high-definition images, even compressed by JPEG, provide much higher accuracy than neural-enhanced ones. 

\tb{Design.}
To resolve the challenge, we make two design choices. 
First, all decisions are made on camera with a lightweight agent fed with global information.
Second, to obtain high accuracy, important frames are delivered in a high-definition format to the edge server rather than the first-compressed-then-enhanced paradigm, like the neural-enhanced method.

\begin{figure}[t]
\centering
\vspace{-1em}
    \includegraphics[width=0.40\textwidth]{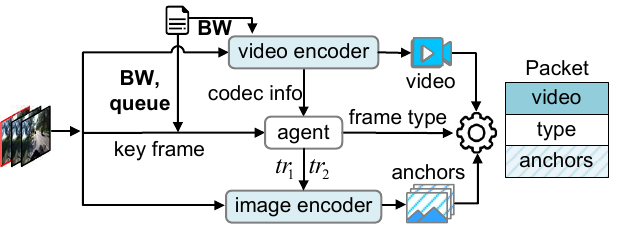}
    \caption{Hybrid encoder.}
    \label{fig:encoder}
     \vspace{-1.3em}
\end{figure}

To this end, we develop an adaptive hybrid encoder consisting of a \emph{video encoder}, an \emph{agent}, and an \emph{image encoder} as shown in Fig. \ref{fig:encoder}. 
The \emph{video encoder} encodes the raw images into a video chunk by adaptively adjusting the bitrate and resolution subject to the bandwidth allocated by the bandwidth controller (see \S \ref{sec:BWcontroller}). 
In particular, the bitrate and resolution are dynamically tuned 
with an adaptive feedback control system \cite{du2020server}, in which different quality configurations (detailed in \S \ref{imple}) are tuned according to the difference between the allocated bandwidth for the current video chunk and that used for the previous chunk.
The \emph{agent}, modelled as deep reinforcement learning, automatically classifies raw images into three types (mapping to three pipelines, detailed in \S \ref{sec:hybriddec}) based on the frame difference information from codec, the video content characteristic from the key (first) frame of each chunk, and the feedback (including the bandwidth allocation and the task queue length on edge) from the edge server.
Importing such feedback is to account for multi-stream server resource contention and its impact on computational latency.
Then, the raw images belonging to the first type, regarded as anchors that offer the highest accuracy improvement, are encoded into HD images via an \emph{image encoder}.
Finally, the hybrid encoder packages and delivers compressed video, frame type information and HD images together to the edge. 

In \S \ref{low_level}, we are going into the details of the frame classify DRL agent. Controlled by the \emph{agent}, 
the number of the anchors
, providing the highest accuracy gain, 
are auto-tuned to make HD images occupy an appropriate bandwidth with the compressed video chunk, so the anchors and video share the allocated bandwidth 
allocated to this stream in an accuracy-latency trade-off. In our evaluation (\S \ref{sec:cw_analysis}), the accuracy-first policy selects sparse anchors, 7\%$\sim$8\%, distributed in one video chunk. 

\subsection{Key Component: Hybrid Decoder \& Execution Pipelines}\label{sec:hybriddec}

\tb{Goal.} 
Edge server in BiSwift responses to execute precise decoding and efficient inference.
However, decoding each frame and naively performing per-frame inference disregards many acceleration and accuracy improvement opportunities.
Thus, we aim to develop a fast executor to speed up end-to-end inference while improving accuracy.
Moreover, this executor must be lightweight enough to support multiple streams.


 
\begin{figure}[t]
\vspace{-1em}
\centering
    \includegraphics[width=0.45\textwidth]{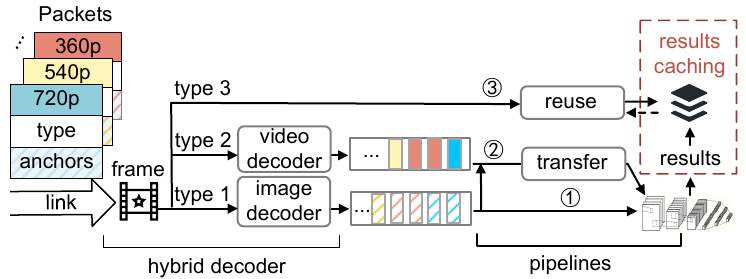}
    \caption{Hybrid decoder.}
    \label{fig:decoder}
    \vspace{-1em}
\end{figure}

\tb{Design.} 
Based on the above goals, we design a hybrid decoder followed by three executing pipelines offering different levels of accuracy and latency (\eg, higher accuracy pipelines exhibiting longer delays) to satisfy a different need from the frame content\cite{AccDecoder}.
When a compressed frame arrives, the hybrid decoder parses the header to identify its type, then severally decodes them by the image or video decoder and delivers them to the corresponding pipelines.
As depicted in Fig. \ref{fig:decoder}, all the type-1 frames (\ie, anchors) are 
HD images, hence decoded with the image decoder, then are fed into the DNN model for inference; the inference results are cached (see pipeline~\ding{172}).
Type-2 frames are decoded by the video decoder and then get enhanced by \emph{quality transfer} from HD images and then fed into DNN to infer and also cache the results (as pipeline~\ding{173}).
To the type-3 frames, pipeline~\ding{174} ``infers'' them by \emph{reuse} module with the cached inference results without decoding. 
%

\begin{figure}[t]
\centering
    \includegraphics[width=0.45\textwidth]{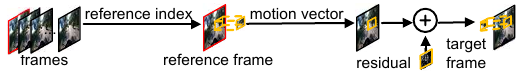}
    \caption{Quality transfer.}
    \label{fig:HRreuse}
     \vspace{-1.5em}
\end{figure}

\eb{Quality transfer.} 
The hybrid decoder enhances the quality and enlarges non-anchor frames 
by leveraging high-quality content from HD images 
with codec information.
Similar to the selective neural-enhancement (\S \ref{sec:background}), this process, as shown in Fig. \ref{fig:HRreuse}, involves: 1) locating reuse blocks on HD images using a reference index, 2) up-scaling the motion vectors (MVs) and interpolating the residual for alignment, 3) combining them into the HD block and seamlessly pasted onto the non-anchor frame. 
However, using such transferring incurs an inevitable accuracy loss.
Fortunately, careful selection of anchors in \emph{ agent} mitigates this issue.

\begin{figure}[h!]
\hspace{-1em}
\centering
\subfigure[Accuracy gain from transfer.] {
 \label{fig:AccTrans}     
\includegraphics[width=0.205\textwidth]{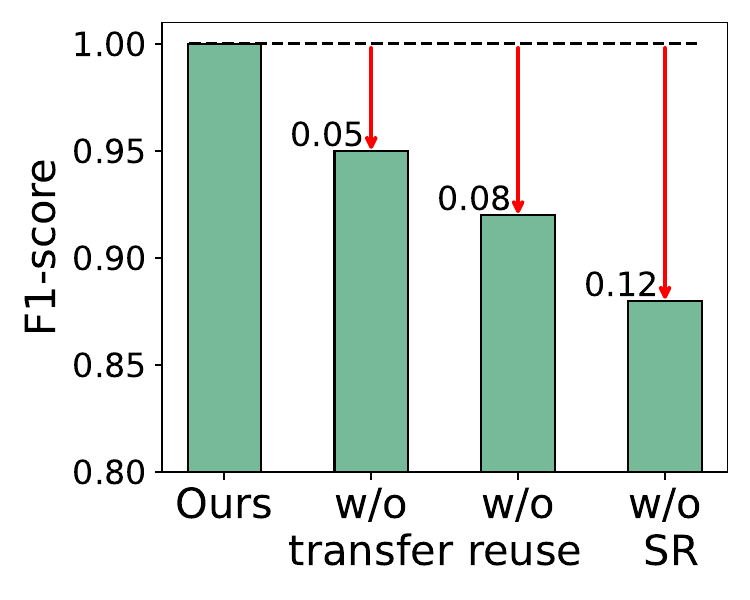}
} 
\hspace{-1.5em}
\subfigure[Time save from reuse.] {
 \label{fig:mvuse}     
\includegraphics[width=0.245\textwidth]{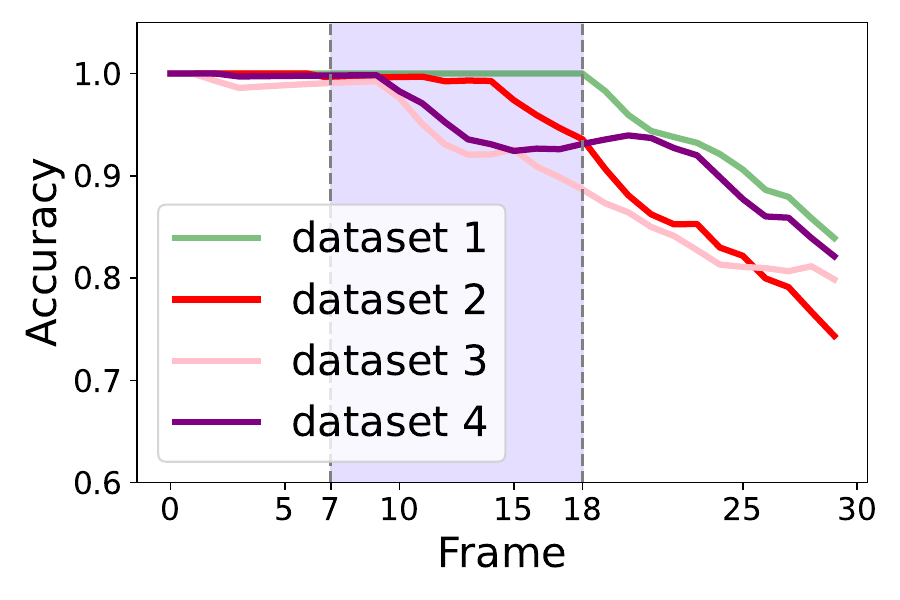}
}
\caption{Performance gain of transfer and reuse module.}
\label{fig:FvR}
\vspace{-1em}
\end{figure}

\eb{Inference results reuse.}
We use object detection, which bounds a box (Bbox) of identifying objects, as an example to illustrate the idea.
Reuse module 1) gets the result of the last inference frame, 2) calculates the mean of all MVs that reside in each Bbox, and 3) shifts each Bbox by the MV mean to the target position.
Our preliminary study in Fig. \ref{fig:mvuse} implies that compared with per-frame inference, reuse brings significant acceleration potential (7$\sim$18 frames in dataset \cite{Yoda_c, Gebhardtdata, Yoda}). 


\subsection{Key Component: Bandwidth Controller}
\label{sec:BWcontroller}
\tb{Problem \& Goal.}
Motivated by the observed poor performance of even bandwidth allocation (see Fig. \ref{fig:band_alloc}), our objective is to develop an analytics-aware bandwidth allocation scheme that ensures fair accuracy across streams.
However, this task is not trivial.
First, this contradicts the agent's goal in the hybrid encoder, which maximizes the performance of the respective data streams by applying as much bandwidth as possible (as discussed in \S \ref{sec:hybridenc}).
Second, the networking environment is dynamic, and the resource requirements of each stream also fluctuate with changes in video content.

\tb{Design.} The edge server, equipped with the bandwidth controller, has a holistic view of all video streams and changing networking resources. Using this information, it dynamically allocates bandwidth resources fairly across the streams based on total available bandwidth, stream-specific details (\eg, content characteristics and inference accuracy), and server state (\eg, queue length of each pipeline).
The optimization problem is formulated as follows:
\begin{equation}\small
\begin{aligned}
\max_{\mathbf{b}} \; &  Fairness(\mathbf{b}, \mathbf{x})\\
\text{s.t.}  
\;\;  & x_c = \arg\max_{x} U_c(x, b_c)\quad \forall c \in C, \\
\;\;  &  L_c(x_c, b_c) \leq \tau \qquad \forall c \in C,\\
\;\;  &  \sum_{c\in C} b_c \leq B \quad\ \forall c \in C,
\end{aligned}
\label{eq:high_level_o}
\end{equation}
where $\mathbf{b}=(b_1,..., b_{|C|})$ represents the bandwidth allocation vector for the stream set $C$; while $\mathbf{x}=(x_1, ..., x_{|C|})$ and $x_c=(x_{c,1}, ..., x_{c,F_c})$, $f\in F_c$ corresponds to the frame classification for stream $c$, where $x_{c,f} \in \{1, 2, 3\}$ indicates the domain of classification of frame $f$ of stream $c$, frame number in a chunk denoted as $F_c$; $L(\cdot)$ denotes the end-to-end latency of a client.
The objective function is to maximize the fairness among streams through optimizing the bandwidth allocation, \ie, given a set of streams $C$, from the global view of the network controller.
The constraints are 1) the optimal frame classification that maximizes the utility $U_c=\mathbb{E}[R_c]$ of each individual stream $c$; 2) the end-to-end analytical latency constraint (within the latency tolerance $\tau$); 3) the available bandwidth constraint (within the available bandwidth $B$).

\tb{Solution.} Regarding the conflicting objectives of frame classification in the hybrid encoder and bandwidth allocation on edge, we approach this optimization problem using a hierarchical structure.
The network resource allocation and frame classification are categorized as high-level and low-level problems, respectively, with the decision variables of the high-level problem influencing the low-level problem.
Such property of dependency inspires us that this problem can be addressed using Bi-Level Optimization (BLO) \cite{liu2021investigating}.
In the next section, we provide a detailed explanation of the system design and the approach to solving this bilevel problem.
\section{Bilevel optimization}\label{sec:bilevel}

\subsection{Low-level: frame classification}
\label{low_level}
\tb{Problem formulation.} 
To offer each stream optimal accuracy online, we formulate the adaptive frame classification as a maximizing the accuracy problem under the latency constraint.
Given a video from camera $c$ containing the set of frames $F_c$, one of the three pipelines is assigned to each frame, which can be expressed as follows:
\begin{equation}
\begin{aligned}
\max_{x_c} \; & \sum_{f\in F_c} Acc(x_{c,f}) \\
\text{s.t.}  \;\;    & \sum_{f \in F_c} L(x_{c,f}|b_c) \leq \tau,
\end{aligned}
\label{eq:model}
\vspace{-1em}
\end{equation}
where
$Acc(\cdot)$ is the accuracy of a frame given bandwidth $b_c$, $L(\cdot)$ is the processing latency of a frame.

\begin{figure}
    \centering
    \includegraphics[width=0.4\textwidth]{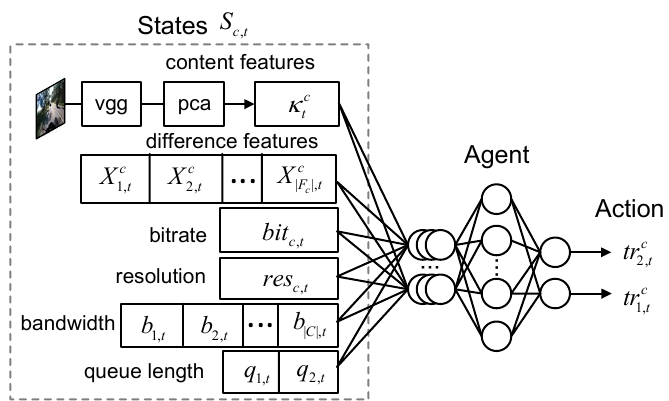}
    \caption{Agents in low level (LN).}
    \label{fig:NN_low}
\end{figure}

\tb{Solution.} 
Such frame classification is not trivial because the optimal frame categorization varies across video sets and differs along with time, even within one chunk \cite{AccDecoder, yeo2020nemo, li2020reducto}.
To this end, we choose DRL to select anchor frames, ``transfer'' frames and ``reuse'' frames (detailed in \S \ref{sec:hybriddec}) in real time while optimizing the overall accuracy.
\begin{itemize}[left=0pt]
    \item
    \tb{Inputs/state}. 
    In Fig. \ref{fig:NN_low}, at chunk $t$, one of BiSwift's learning agents $c$ (where each camera has an agent) takes state inputs $S_{c,t}=(\boldsymbol{\kappa_{t}^c},\mathbf{X_{t}^{c}},bit_{c,t},res_{c,t},\mathbf{b_{t}},\mathbf{q_t})$ to its neural networks. Here, $\boldsymbol{\kappa_{t}^c}=(\kappa_{1,t}^{c}, \kappa_{2,t}^{c},..., \kappa_{128,t}^{c})$ represents the content feature of the key frame, and $\mathbf{X_{t}^{c}}=(X_{1,t}^{c}, X_{2,t}^{c},..., X_{|F_c|,t}^{c})$ with
    $X_{f,t}^{c}$ being the difference feature between frame $f$ and the last inference frame before $f$ \cite{AccDecoder}. 
    Additionally, based on the allocated bandwidth for each camera, the agent can adaptively choose the appropriate video bitrate $bit_{c,t}$ and resolution $res_{c,t}$ using an adaptive feedback control system (\S \ref{sec:hybridenc})
    . The vector $\mathbf{b_{t}}=(b_{1,t}, b_{2,t}, ..., b_{|C|,t})$ represents the allocated bandwidth for each camera. The queue length of the inference pipelines~\ding{172}~\ding{173} in the server is denoted by $\mathbf{q_t}=(q_{1,t},q_{2,t})$, providing insights into the GPU payload on edge resulting from previous frame classification.
    \item
    \tb{Action.}
   The action involves two thresholds, $tr_{1,t}^c$ and $tr_{2,t}^c$ for chunk $t$ of camera $c$. $tr_{1,t}^c$ helps to select anchor frames which transfer their quality to the remaining frames for scale-up, and $tr_{2,t}^c$ helps to select inference frames followed by DNN models such as object detection. The remaining frames leverage the results of the inference process by utilizing frame references \cite{AccDecoder}. Given the two thresholds, the frames are classified into three types as follows:
    \begin{equation}
     x_{c,f}=\begin{cases}
            1& \text{ if }   X_{f,t}^c>tr_{1,t}^c,\\
            2& \text{ if }   X_{f,t}^c\leq tr_{1,t}^c  \text{ and } R_{f,t}^c>tr_{2,t}^c,\\ 
            3 & \text{ others.} 
        \end{cases}
    \label{tr_to_type}
    \end{equation}
where $R_{f,t}^c$ is the accumulated residual from the last inference frame to frame $f$ for camera $c$ at chunk $t$.  

  \item
    \tb{Reward.} The reward is designed to include two aspects \cite{AccDecoder}, namely, the average accuracy of chunk $t$ and the latency penalty when timeout. $\alpha_1$ and $\alpha_2$ are the weight factors for balance. For simplicity, we omit the subscript $c$ (camera index) in the following, except for.
    \begin{equation}
    r_{c,t}= \frac{\alpha_1}{|F_t|} \sum_{f\in F_t} Acc_f(tr_{1,t}, tr_{2,t})-\alpha_2 P_t(tr_{1,t}, tr_{2,t}),
    \label{low_reward}
    \end{equation}
$P_t$ is a penalty function of chunk $t$ for latency exceeding the tolerance $\tau$.
When latency $L(tr_{1,t}, tr_{2,t})>\tau$, $P_t(tr_{1,t}, tr_{2,t})=1$; otherwise, $P_t(tr_{1,t}, tr_{2,t})=0$.
In which, $ L(tr_{1,t}, tr_{2,t})= time_{trans}^{t}+time_{queue}^{t}+time_{comp}^{t}$ is the entire time of transmission, queuing, and pipelines executing.
    
\end{itemize}
\subsection{High-level: analytics-aware bandwidth allocation}\label{high_level}
\tb{Problem formulation.} 
The goal of the high-level is to optimally allocate bandwidth for each stream to maximize fairness, so we concretize the object function in Eq. \ref{eq:high_level_o} as follows. For the sake of formula simplicity, we omit the
subscript $t$ (video chunk index) in the variables in the following
formula.
\begin{equation}
\begin{aligned}
\max_{\mathbf{b}} \; &  \min_{c \in C} r_c(b_c, tr_1^c, tr_2^c, S_{high}) \\ 
\text{s.t.}
 \;\;  & \sum_{c \in C} b_c \leq B, \; b_c \geq 0, \forall c \in C,\\
 \;\;  & tr_1^c, tr_2^c = LN_c(S_{c})
\end{aligned}
\label{eq:high_level}
\end{equation}
where $LN_c$ refers to the low-level network responsible for frame classification of camera $c$, as depicted in Fig. \ref{fig:NN_low}. The state of the low-level network of camera $c$,
denoted as $S_{c}$, is detailed in \S \ref{low_level}. Moreover, $r_c$ is reward of camera $c$ and $S_{high}$ is the observed state from the environment.


\tb{Solution.} 
The bandwidth controller 1) adaptively adjusts bandwidth allocation when the video content or networking environment changes over time. 
2) is equipped with the ability to handle the non-convexity of DNN accuracy, which is always difficult in finding the optimal solution to a problem because it may become trapped in local optima. However, this property is often overlooked by prior heuristic algorithms \cite{sheng2022VA}. 3) can search for the optimal solution in a vast exponential search space (see details in \S \ref{comp}).
Therefore, we adopt reinforcement learning to solve this bilevel problem gracefully. 
We omit the details of joint training due to limited space.

\begin{figure}[t]
    \centering
    \includegraphics[width=0.25\textwidth]
    {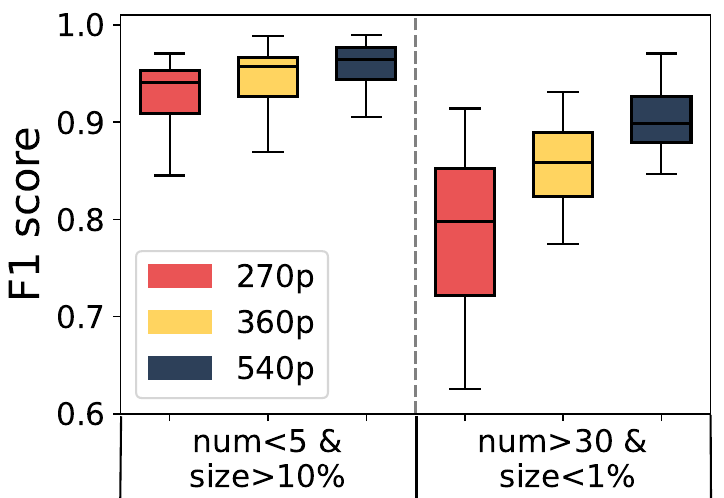}
    \captionsetup{justification=raggedright}
    \caption{Difference accuracy robustness
of the videos with different object numbers and sizes.}
    \label{fig:diff_obnum}
\end{figure}

\begin{itemize}[left=0pt]
\item{
\tb{Inputs/state.} BiSwift's bandwidth controller takes 
${S_{high}}=\left(\mathbf{num},\mathbf{size},\mathbf{r},\mathbf{b_{L}},\mathbf{acc},\mathbf{p}\right)$ 
as states. The $S_{high}$ comprises the following components. The first component includes 
$\mathbf{num}=(num_1,num_2, ..., num_{|C|})$ (in which $num_c$ means the average object number of camera $c$) and $\mathbf{size}=(size_1,size_2, ..., size_{|C|})$ (in which $size_c$ means the average object size of camera $c$). The videos with different object numbers and sizes always exhibit varying detection difficulty influenced by factors such as object density and camera distance (as mentioned in \S \ref{moti_3}). 
As shown in Fig. \ref{fig:diff_obnum}, the video with rare objects presents better accuracy robustness under different resolutions.
Due to the temporal similarity in video content, $\mathbf{num}$ and $\mathbf{size}$ at chunk $t$ can be estimated from the object detection results of the previous chunk $t-1$. In addition, $\mathbf{r}=(r_1, r_2, ..., r_{|C|})$, in which $r_c$ is the average residual of camera $c$ at the current chunk, is designed as one of the DRL inputs. The video with larger residual always includes more I frames (Intra-coded frame)\cite{wiegand2003overview}, indicating faster moving speed or more frequent scene switching. 
The next component of $S_{high}$ reflects the previous bandwidth orchestration configuration and resulting gains, \ie, $\mathbf{b_{L}}$ and $\mathbf{acc}$.
The last component $\mathbf{p}$   
indicates the anchor proportion of each stream in pipeline~\ding{172} on edge,
reflecting the low-level decision of all cameras.
 }
\item{
\tb{Action.} The action associated with the high-level problem involves determining the proportion of bandwidth for each camera, denoted as $\mathbf{b}=(b_1, b_2, ..., b_{|C|})$.
}
\item{\tb{Reward.} Given that BiSwift aims to maximize the reward of all streams fairly, described in \eqref{high_level}, the high-level reward takes the minimum reward of all streams as its reward.
\begin{equation}
    r_{high}=\min_{c \in C} r_c
\end{equation}
where $r_c$ is reward of camera $c$. 
}      
\end{itemize}  
\subsection{Computational Complexity} 
\label{comp}
Firstly, the searching space of the optimal selection space is $\mathcal{O}(3^k \xi^N)$, 
$k$ is the frame number of streams per second, $\xi$ represents the number of bandwidth levels after discretization, and $N$ is the number of cameras. 
The searching space of BiSwift is $\mathcal{O}(|a| +|a'|)$, where $|a|$ and $|a'|$ are the number of possible actions for the agent and the bandwidth controller. The complexity of BiSwift is significantly lower than that of searching the optimal selection space. We omit the details due to limited space.
\section{Implementation}

\subsection{System Settings, Dataset, and Baselines}
\label{imple}
\tb{System settings}.
Components on edge, including bandwidth controller and pipelines, run on an Ubuntu 18.04 instance (5.4.0-113-generic) with Intel(R) Xeon(R) Gold 6226R CPU at 2.90GHz and 1 NVIDIA GeForce RTX 3070(8GB) GPU. 
We use the H.264 codec \cite{libvpx}, JM 19.0 version open source code \cite{JM}, with 2470 LoC (Lines of Code) changes. 
Moreover, we use NVIDIA nvJPEG, a GPU-accelerated JPEG codec library, to encode images. 
For the cameras, multiple laptop terminals are utilized to simulate the cameras, performing image capture and data transmission based on DDS's \cite{du2020server} simulation mode. 
Furthermore, the overall available bandwidth given by an FCC broadband network trace \cite{FCC} and the allocated bandwidth for each terminal are set using WonderShaper \cite{wondershaper}.

\tb{Dataset.}
Our video dataset is gathered from various publicly available real-time surveillance camera streams deployed around the world. 
We collected video clips from different camera sources, resulting in a dataset with diverse properties such as illumination, object density, road type, and direction. 
All videos can be obtained from YouTube \cite{Gebhardtdata, jackson, auburn2} and Yoda \cite{Yoda_c, Yoda_m, Yoda}, captured at 30 fps.  We set five levels of video quality configuration: $bitrate \in \{500,1000,1500,2000,5000\}$ and their corresponding $resolution \in\{270p,360p,540,720p,1080p\}$.


\tb{Baselines.}
%
We compare BiSwift's performance with the following baselines:
(1) AccDecoder \cite{AccDecoder} applies DRL on adaptive frame classification for DNN-based inference and uses a super-resolution deep neural network to enhance video quality. 
(2) Reducto \cite{li2020reducto} uses a dynamic threshold learned by the server to filter out unnecessary frames on cameras to save bandwidth. 
(3) NeuroScaler \cite{yeo2022neuroscaler} enables selective enhancement in live neural-enhanced streaming by using super-resolution on anchor frames. 
Naive NeuroScaler is designed for the quality of experience (QoE) improvement, so we extend it for video analytics, named NeuroScaler*, by replicating its frame selection algorithm but adjusting the number of anchor frames under latency constraints; 
then, run object detection on anchor frames and reused the results on others.
To make the above methods suit multi-stream, we apply even bandwidth allocation to them. 
\setlength{\itemsep}{0pt}
\setlength{\parsep}{0pt}
\setlength{\parskip}{0pt}
\subsection{DNN Implementation}
\setlength{\itemsep}{0pt}
\setlength{\parsep}{0pt}
\setlength{\parskip}{0pt}
\tb{DRL settings}.
As low level, we leverage an actor-critic network as our agent by setting $\alpha_1=\alpha_2=0.5$ and $\tau=1s$ in \eqref{low_reward} for each chunk. 
The learning rates of the Adam optimizer of the actor and the critic network are 0.005 and 0.01, respectively; while the discount factor of the reward $\gamma$ is set to 0.9. 
We use a two-layer MLP with 128 units to implement the policy networks and another two-layer MLP with 128 units to implement the value networks in DRL.
All neural networks used employ ReLU as activation functions.
As for the high-level DRL network, we use the soft actor-critic algorithm \cite{haarnoja2018soft}. 
The learning rates for the policy network, value network, and Q function networks are 0.001, 0.003, and 0.0003 respectively; while the target update rate $tau$ and the discount factor are separately set as 0.02 and 0.9. The size of the replay buffer is $10^4$ and we take a minibatch of $128$ to update the network parameters. Unlike the low-level agent, the policy networks here use a four-layer MLP with 256 units and the value networks use a three-layer MLP with 256 units. The high- and low-level DRL are trained jointly, exchanging experience data or decision results as described in \S \ref{sec:bilevel}.

\label{train}
\tb{Inference DNN settings}.
To evaluate the performance of BiSwift, we focus on its use for object detection. 
Specifically, we tested two object detection architectures including Faster R-CNN \cite{ren2015faster} and YOLOv5 \cite{yolo}. 
Both models were pre-trained on the COCO dataset \cite{COCO}. 

\section{Evaluation}

\subsection{Overall performance of BiSwift}
\noindent


\begin{figure}[h!]
\setlength{\abovedisplayskip}{0pt}
\setlength{\belowdisplayskip}{0pt}
\centering
\subfigure[Processing throughput.] {
 \label{fig:baseline_throu}     
\includegraphics[width=0.22\textwidth]{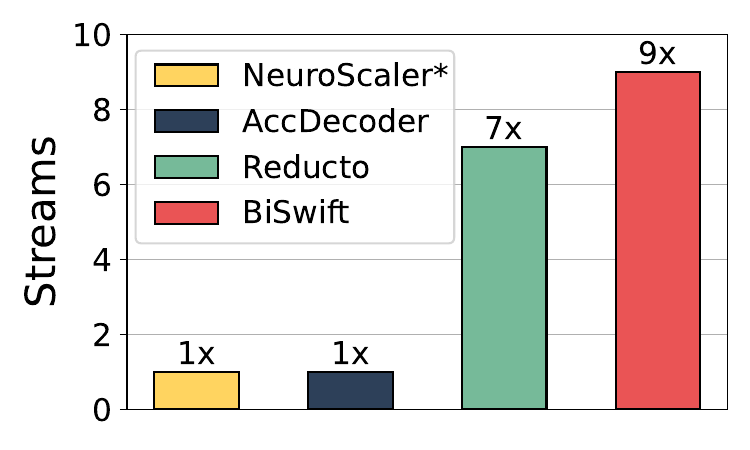}
}
\subfigure[Bandwidth cost.] {
 \label{fig:baseline_band}     
\includegraphics[width=0.22\textwidth]{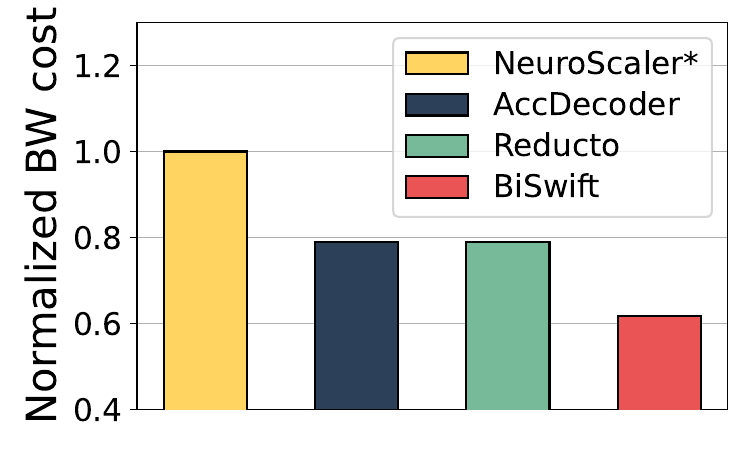}
}
\subfigure[Latency and inference accuracy.] {
 \label{fig:baseline_acc}
\includegraphics[width=0.4\textwidth]{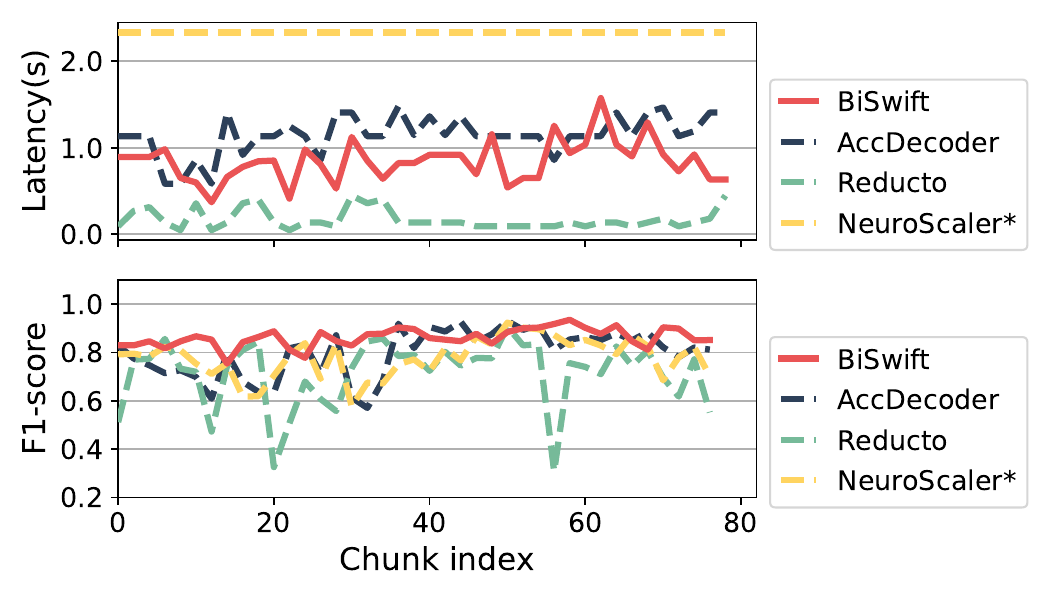}
}
\caption{End-to-end performance against baselines.}
\label{fig:baselines}
\end{figure}
\noindent
\tb{Throughput improvement and bandwidth cost decrease.} Fig. \ref{fig:baseline_throu} illustrates the end-to-end average throughput. BiSwift significantly improves the end-to-end processing throughput by 9$\times$ compared with AccDecoder and NeuroScaler*. Reducto can also achieve a higher throughput but still lower than BiSwift, whereas other baselines offer a lower throughput. Both AccDecoder and NeuroScaler* employ super-resolution techniques, selecting as many anchor frames as possible to improve accuracy while meeting real-time constraints. However, this results in lower throughput due to the increased computational requirements. In terms of bandwidth usage as shown in Fig. \ref{fig:baseline_band}, BiSwift incurs the lowest bandwidth cost by using an adaptive hybrid encoder, allowing for more aggressive video compression, because low-level pipeline selection algorithm \ref{low_level} can adapt anchor frames to compensate for the compressed video's decreased accuracy. NeuroScaler* also employs hybrid encoding and transmits additional anchor frames, it results in higher bandwidth usage. \\
\noindent
\tb{Accuracy and latency.} We demonstrate that BiSwift can effectively improve accuracy-latency trade-off via adaptive pipeline assignment.
Fig. \ref{fig:baseline_acc} show the per-chunk of average latency and accuracy (\ie, f1-score as a measure of accuracy) of all streams. BiSwift achieves higher and more stable accuracy ($\sim$0.87) compared with other algorithms (AccDecoder $\sim$0.79,  NeuroScaler* $\sim$0.77 and Reducto $\sim$0.72). The accuracy of other algorithms for multiple streams fluctuates significantly between the 15th to 25th chunk, with severe degradation. This is due to the larger difference of streams with the object number from 0$\sim$5 (in stream 1 and 2) to $>$30 (in stream 3 and 4). BiSwift's bandwidth controller, which allocates bandwidth based on the characteristics of each stream, assign more bandwidth to stream 3 and 4 to achieve higher accuracy.

\begin{figure}
    \centering   \includegraphics[width=0.25\textwidth]{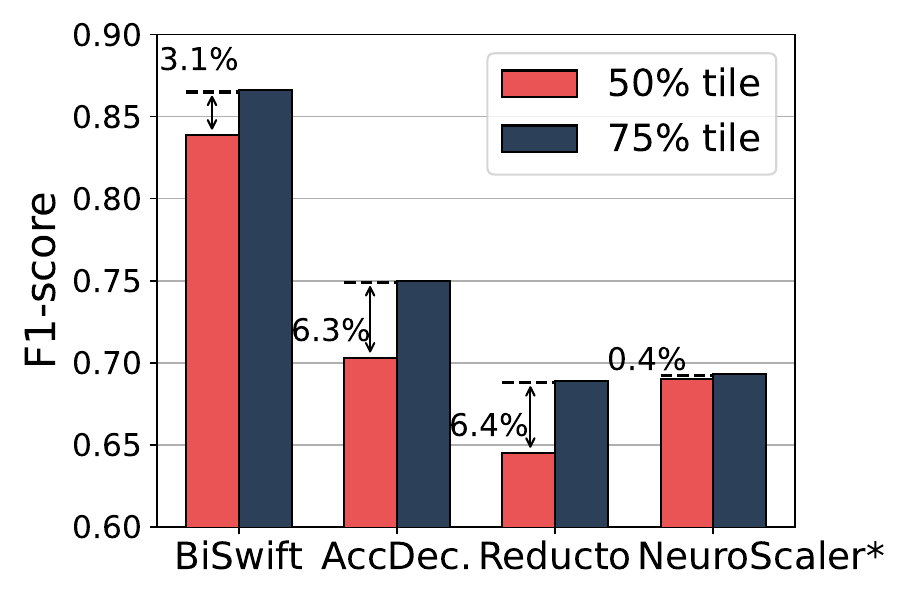}
    \caption{Accuracy distribution and fairness.}
    \label{fig:baseline_tile}
\end{figure}

\noindent
\tb{Fairness across multi-stream.} We conduct a study on the accuracy of all streams over a period of time and sorted the data in ascending order. We select the  50\% and 75\% percentiles and observe that BiSwift not only has a high average accuracy ($\sim$85\%), but also has relatively small difference in accuracy (3.2\%)  across  streams at different percentiles as shown in Fig. \ref{fig:baseline_tile}. This indicates that the fairness metric in \S \ref{high_level} is effective and BiSwift's performance is stable across heterogeneous streams. 

\subsection{Component-wise Analysis} \label{sec:cw_analysis}
\begin{figure}[h!]
\centering
\hspace{-0.7em}
\subfigure[Accuracy breakdown.] {
 \label{fig:break-a}     
\includegraphics[width=0.22\textwidth]{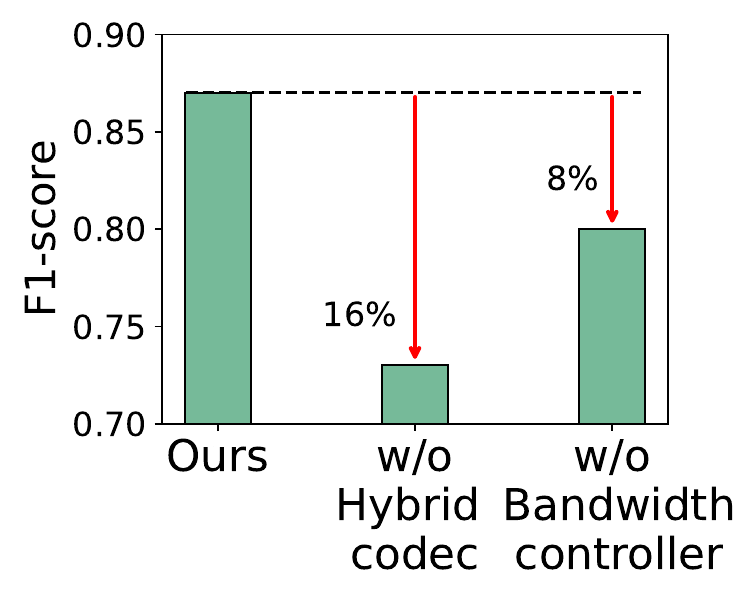}
}
\subfigure[Latency breakdown.] {
 \label{fig:break-b}
\includegraphics[width=0.23\textwidth]{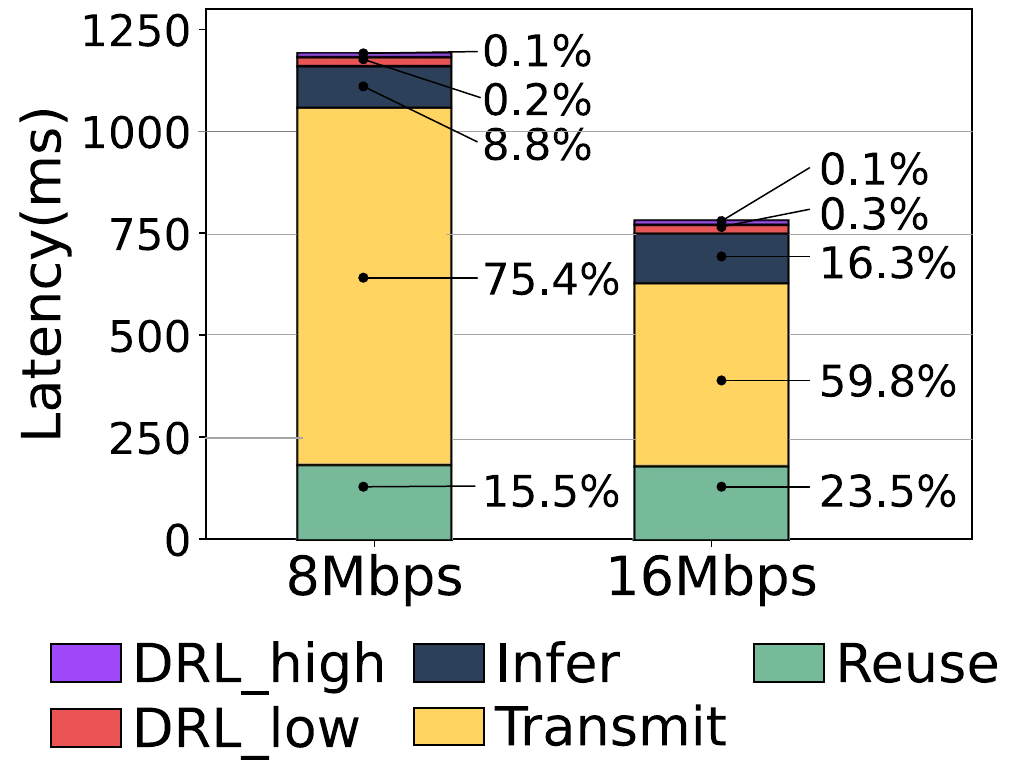}
}
\caption{Performance breakdown on accuracy and latency.}
\label{fig:break}
 \end{figure}

\tb{Accuracy breakdown.} As shown in Fig. \ref{fig:break}, we break down the accuracy and the latency of BiSwift's two main components: adaptive hybrid encoder and bandwidth controller. 
Fig. \ref{fig:break-a} breaks down the accuracy of the adaptive hybrid encoder and bandwidth controller. 
Without adaptive hybrid encoder but uniform anchor selection leads to a decrease in accuracy of 16\%;
while replacing the bandwidth controller with even bandwidth allocation results in a decrease in accuracy of 8\%.

\tb{Latency breakdown.} Fig. \ref{fig:break-b} shows the latency breakdown in processing each chunk, where we use a wireless link over 8Mbps and 16Mbps. 
Most of the latency of BiSwift is attributed to the transmission of data packets ($\sim$67\%), consisting of compressed video chunks and high-definition HD anchors. 
When the available bandwidth is larger, the transmission time decreases significantly by 48\%. 
Second, although the time cost of reuse per frame ($\sim$6ms) is small, the time cost of reuse occupies $\sim$19\% of the latency in total due to more than 90\% frames belonging
to pipeline~\ding{174}. 
Third, 
8\% of the frames undergo inference, with 6\% of the frames entering pipeline~\ding{172} (inference) and 2\% of the frames entering pipeline~\ding{173} (transfer+inference), so the latency of inference is $\sim$13\%.
Compared with other components, the lightweight DRL models incur a $<$10ms inference time with the low-level (\ref{low_level}) DRL network 7ms on average and the high-level (\ref{high_level}) DRL network 2ms on average, occupying an extremely small proportion. 
Besides, the interval of execution of BiSwift's bandwidth controller (high-level DRL) can be adjusted (every 10s in our experiment), with time being negligible.
 
\tb{Computional cost.} BiSwift's transmission of data and the computation on the server can be parallelized in a pipelined manner, resulting in the actual server computation latency being negligible. As a result, a single GPU is capable of processing up to 8 video streams.
The computational cost of the proposed system was evaluated using NVIDIA Nsight \cite{nsight}. The GPU memory usage was 2$\sim$3GB (8GB in total) and GPU  achieve close to 100\% utilization for 40\%$\sim$70\% of the time. 
Given that the computational demand of DRL models and reuse is low, we execute them on CPU. The CPU utilization ranges from 45\% to 50\%, with utilization of 30\%.

\begin{figure}[h!]
\centering
\hspace{-1em}
    \subfigure[highways] {
     \label{fig:bub-b}     
    \includegraphics[width=0.235\textwidth]{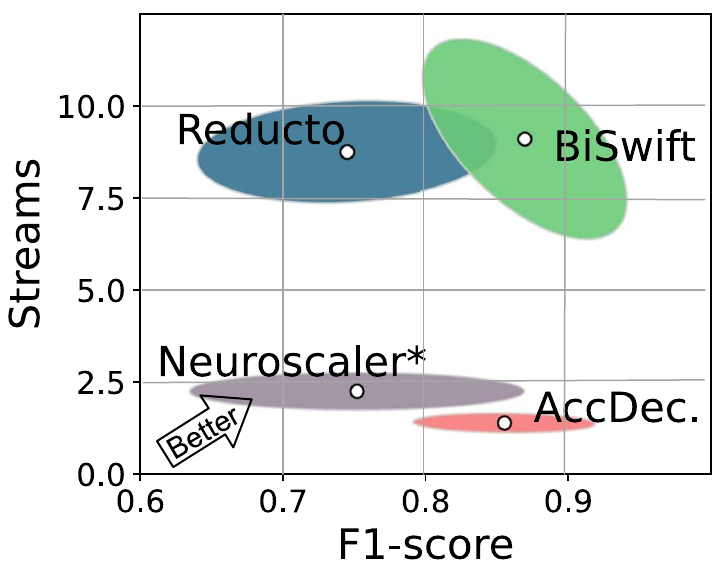}
    } 
    \subfigure[crossroad] {
     \label{fig:bub-d}     
    \includegraphics[width=0.235\textwidth]{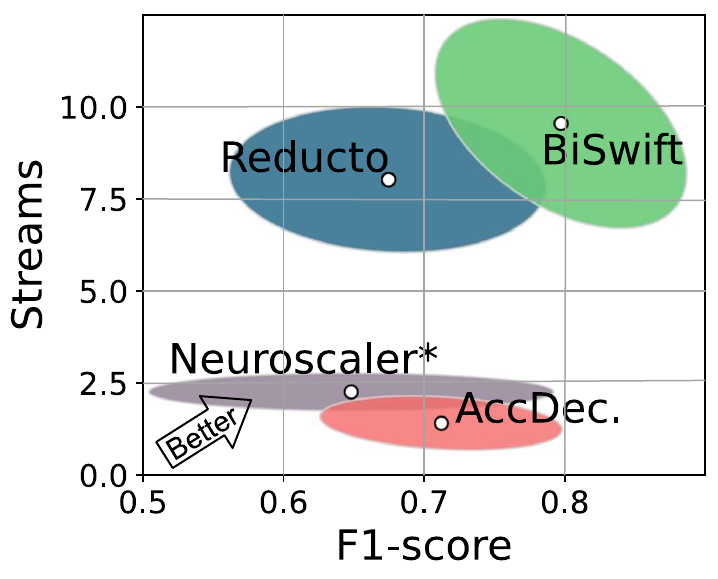}
    }
    \hspace{-1em}
    \captionsetup{justification=raggedright}
    \caption{BiSwift vs. baselines in terms of the throughput and inference accuracy on various video datasets using the YOLOv5 model.}
    \label{fig:bubbles}
\end{figure}

\subsection{BiSwift vs. Existing VAPs}
 We show the advantage of BiSwift in accuracy and throughput compared with the baselines.  Fig. \ref{fig:bubbles} compares the performance distribution of BiSwift with the baseline over the YOLOv5 model and various types of videos (\ie, highways and crossroads).  BiSwift achieves 5-17\% higher inference accuracy than the baselines and 10-89\% more processing streams than the baselines except for Reducto.

\section{Related Work}
\noindent
\textbf{Video analytics pipeline}.
Mainstream video analytics pipelines primarily focus on two objectives: bandwidth saving and improvement of inference accuracy. 
To achieve bandwidth saving, cameras can leverage their computing power for analytics 
\cite{liu2019edge, Kang2017NoScope,chen2015glimpse} or use cloud feedback to discard frames with irrelevant information \cite{jiang2018chameleon, zhang2018awstream, du2020server}, thus conserving bandwidth. 
Additionally, selectively streaming regions of interest within frames can further save bandwidth in video streaming systems \cite{zhang2015design, Marwa20region, hsieh2018focus}. 
To improve inference accuracy, researchers draw inspiration from the success of image enhancement in video streaming \cite{yeo2018neural, kim2020neural, yeo2020nemo, dasari2020streaming, yi2020supremo} and explore its application in machine-centric video analytics \cite{wang2022enabling, yi2020eagleeye,wang2019bridging}. 
Techniques like super-resolution and GAN-based image enhancement have shown promising results in increasing inference accuracy. 
Scaling video analytics has garnered attention, with efforts in cross-camera correlation \cite{jain2020spatula, canel2019scaling} to speed up VAP configuration and resource management for efficient computing \cite{cui2021enable, padmanabhan2022gemel} and bandwidth usage \cite{Yuan2022VSiM}.

\tb{DRL and Bi-level RL}.
DRL is well suited to tackle problems requiring longer-term planning using high-dimensional observations, which is the case of both dynamic pipeline selection and bandwidth slicing.
Bi-level RL \cite{liu2021investigating} refers to a class of RL algorithms where there are two levels of optimization: an upper-level that aims to optimize the overall performance of agents, and a lower-level that optimizes the parameters of the policies.
Bi-level RL has gained significant attention in recent years due to its applications in various domains.


\section{Conclusion and Discussion}
In this paper, we looked at the scalability issues in video analytics.
We propose BiSwift, a scalable framework for live video analytics that addresses the challenges with a holistic approach.
BiSwift enables access to high-definition images for high accuracy by proposing the hybrid encoder; develops a frame classifier (\ie, a DRL agent) to allocate bandwidth for image (anchors) and video delivery; designs a global analytics-aware bandwidth management to optimize the accuracy fairness. 
In our evaluation on the edge server under real-world network trace, BiSwift can deliver nearly an order of magnitude improvement in scalability.

\section{Acknowledge}
This work was supported in part by the National Natural Science Foundation of China under Grant 62272223, U22A2031, 61832005, in part by the Collaborative Innovation Center of Novel Software Technology and Industrialization, Nanjing University, in part by the Jiangsu High-level Innovation and Entrepreneurship (Shuangchuang) Program, in part by Tsinghua University (AIR) - AsiaInfo Technologies (China), Inc. Joint Research Center (Grant 20203910074), in part by Shuimu Tsinghua Scholar Program (Grant 2023SM201) and in part by EU Horizon CODECO projects (Grant 101092696).

\bibliographystyle{IEEEtran}
\bibliography{ref}

\newpage

\end{document}